\documentclass{article}
\usepackage{spconf}
\usepackage{amsmath}
\usepackage{graphicx}
\usepackage[
    hidelinks,
    pdftitle={Input Representation, Not Model Size Or Pretraining Scale, Shapes Speaker Verification Robustness To Temporal Degradation}, 
    pdfauthor={Erdem Baha Topbas, Srikanth Madikeri, Elisa Pellegrino, Volker Dellwo},
]{hyperref}
\usepackage{booktabs}
\usepackage{siunitx}
\usepackage[capitalise]{cleveref}
\usepackage{mathptmx}
\usepackage{cite}
\usepackage{eso-pic}
\usepackage[dvipsnames]{xcolor}

\DeclareSIUnit\pct{\char`\%}

\definecolor{EP}{HTML}{D55E00}

\ninept

\title{INPUT REPRESENTATION, NOT MODEL SIZE OR PRETRAINING SCALE, SHAPES SPEAKER VERIFICATION ROBUSTNESS TO TEMPORAL DEGRADATION} 

\name{Erdem Baha Topbas\qquad Srikanth Madikeri\qquad Elisa Pellegrino\qquad Volker Dellwo}
\address{Department of Computational Linguistics, University of Zurich\\
\href{mailto:erdembaha.topbas@uzh.ch}{\url{erdembaha.topbas@uzh.ch}}}

\begin{document}

\maketitle

\AddToShipoutPictureBG*{%
  \AtPageLowerLeft{%
    \hspace{0.75in}%
    \raisebox{0.6in}{%
      \parbox{6.9in}{\footnotesize\color[gray]{0.3} This work has been submitted to the IEEE for
      possible publication. Copyright may be transferred without notice, after
      which this version may no longer be accessible.}%
    }%
  }%
}

\begin{abstract}

Speech remains intelligible to listeners under heavy temporal degradation, but whether speaker identity survives is largely unexplored. We evaluate six automatic speaker verification (ASV) models, trained on temporally intact speech, and tested under three degradations that remove different combinations of temporal cues. The models form three pairs, one per architecture family (convolutional, transformer, conformer), each pairing a waveform-input with a mel filterbank-input model. All models degrade, but robustness tracks input representation rather than pretraining scale or model size. In all nine family-by-condition comparisons the filterbank-input model degrades less, with eight of nine \qty{95}{\percent} confidence intervals excluding zero, even when it is smaller or has less pretraining than its waveform counterpart. In the controlled convolutional pair, where the models differ only in input representation, the filterbank-input model's equal error rate rises from \qty{1.1}{\percent} to \qty{17.1}{\percent} under the heaviest tested degradation, against \qty{2.9}{\percent} to \qty{29.7}{\percent} for its waveform counterpart.


\end{abstract}


\section{Introduction}
\label{sec:intro}

\begin{figure}[t]
    \centering
    \includegraphics[width=\linewidth]{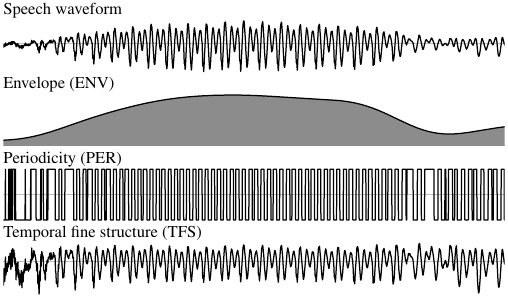}
    \caption{
    Temporal decomposition in a \qty{15}{\milli\second} window of an utterance from the VoxCeleb1 corpus~\cite{Nagrani2017}.}
    \label{fig:temporal-decomposition}
\end{figure}

Temporal information in speech can be decomposed into three components: envelope, periodicity, and temporal fine structure, each with different dominant fluctuation rates~\cite{Rosen1992} as illustrated in \cref{fig:temporal-decomposition}. The envelope (ENV) captures amplitude modulations and energy, corresponding to the slowest fluctuations in a waveform (\qtyrange{2}{50}{\hertz}). The periodicity (PER) corresponds to the medium-rate fluctuations (\qtyrange{50}{500}{\hertz}) and contains zero-crossing related information. Temporal Fine Structure (TFS) contains the fastest temporal detail (\qty{500}{\hertz} to \qty{10}{\kilo\hertz}). These three components characterize temporal information in a speech waveform.

Each component encodes segmental and prosodic cues that contribute to speech intelligibility, such that speech remains sufficiently comprehensible even when one component is removed. For example, infinite peak clipping (IPC)~\cite{Licklider1948}  replaces each sample value with its sign, while preserving speech intelligibility, with listener accuracy between \qty{50}{\percent} and \qty{90}{\percent}. In terms of temporal decomposition, this corresponds to retaining PER while discarding ENV and TFS. 
Near-perfect speech intelligibility has been observed in the condition of substantially reduced spectral information, when temporal envelopes extracted from a limited number of frequency bands were used to modulate band-limited noise~\cite{Shannon1995}.
In conditions of contrasting speech information from ENV and TFS, \cite{Smith2002} shows that ENV information guides word recognition.

Although previous research shows that speech intelligibility is preserved under temporal degradation~\cite{Licklider1948,Smith2002,Shannon1995}, the extent to which speaker-specific information is retained in degraded speech and supports speaker recognition remains largely unexplored.
Our research expands recent work investigating the robustness of ASV systems under varied acoustic conditions~\cite{Baali2025, Nandwana2019} by testing the robustness of speaker verification models under heavy temporal degradation that remains intelligible to humans.
We evaluated six widely used ASV models across two input representations, waveforms and mel filterbanks~\cite{Davis1980}, and three architecture families: convolutional~\cite{He2016}, transformer~\cite{Vaswani2017}, and conformer~\cite{Gulati2020}. All models are trained on unmodified speech and evaluated on three degraded conditions: PER retains periodicity alone, PER+ENV restores the amplitude envelope, and PER+TFS restores temporal fine structure. This design allows us to examine how architecture family, parameter count, pretraining, and input representation affect ASV performance under temporal degradation. On VoxCeleb1-O~\cite{Nagrani2017} we find two results. Filterbank-input models degrade less than their waveform-input counterparts in all nine architecture family $\times$ condition comparisons; in the controlled convolutional pair, which differ only in input representation, EER in the PER condition is \qty{17.1}{\percent} against \qty{29.7}{\percent}. Restoring temporal fine structure recovers a mean of \num{10.9} percentage points of EER, against \num{1.3} for restoring the envelope, an ordering that holds in all six models.
\section{Methodology}
\label{sec:method}

\subsection{Temporal Degradation Techniques}
\label{subsec:tempo-decomp}

The temporal components of speech can all be acquired from a time-domain waveform $x[n]$ with $n\in\left\{0,\dots,N-1\right\}$, where $N$ is the number of samples in the waveform.


The \textbf{PER} condition is acquired by IPC, defined below.

\begin{equation}
\label{eq:ipc}
    \text{IPC}[n] = \begin{cases}
        +1, & x[n]\geq 0\\
        -1, & x[n]< 0
    \end{cases}
\end{equation}

To match~\cite{Licklider1948}, no voice activity detection or silence-trimming was used. 


\textbf{PER+ENV} is obtained by following~\cite{Drullman1994}. Hilbert envelope~\cite{Oppenheim1999} of the waveform undergoes a fourth-order Butterworth low-pass filter at \qty{30}{\hertz}, applied forward--backward for zero phase distortion (effective eighth-order response, \qty{-3}{\decibel} at \qty{26.8}{\hertz}) to remove high-frequency temporal fine structure. Then an element-wise floor is applied to prevent zeros in this filtered envelope, by setting a lower threshold at \num{0.01} times the envelope maximum. This final smooth envelope is all positive, and to reflect the expected shape of a waveform, it is multiplied by the IPC signal.
Unlike in~\cite{Smith2002}, we apply the decomposition to the broadband signal rather than to individual bands. This is motivated by the fact that the systems tested here operate on broadband input, and mirroring the perceptual methodology would not be appropriate here. Because the components derived by our methods are based on broadband and not narrow-band signals, the magnitudes reported here cannot be compared directly with results from studies that primarily use narrow-band components.

\textbf{PER+TFS} can be acquired by
\begin{equation}
    \text{PER+TFS}[n]=\frac{x[n]}{\text{ENV[n]}},
\end{equation}
where ENV represents the Hilbert envelope. No floor is applied to the envelope in this step. Since the analytic magnitude satisfies $\text{ENV}[n]\geq\left|x[n]\right|$ by construction, the quotient is bounded in $[-1,1]$ at every sample however small the envelope becomes. The operation divides out the amplitude contour while retaining the oscillations of the waveform. PER and PER+TFS share an identical zero-crossing pattern and differ only in the waveform shape between crossings. The contrast between them therefore isolates fine structure in the amplitude domain, independent of crossing timing.

No canonical method exists for removing periodicity while leaving the other components intact. Therefore, only periodicity retaining conditions are included in the evaluation.

\subsection{Data and Corpora}
\label{subsec:data-corpora}
For training, we used the VoxCeleb2 corpus~\cite{Chung2018}, using \qty{2}{\second} random crops~\cite{Wang2023}.
Training data was augmented using additive noise from the MUSAN corpus~\cite{Snyder2015} and reverberation added through convolution~\cite{Ko2017}.
Augments were added with a probability of \qty{60}{\percent}, where the signal-to-noise ratio of the additive noise was randomly selected to be between \qty{0}{\decibel} and \qty{20}{\decibel}. The models were trained exclusively on speech with intact temporal cues, which we will henceforth refer to as \textit{unmodified}.
For evaluations, we used the VoxCeleb1-O trial pairs~\cite{Nagrani2017}.
All verification pairs were presented condition-matched.



%

We selected the following models based on their widespread use and reported performance on speaker verification, prioritizing architecture families for which counterparts using different input representations were available. The four pretrained models we chose are \textbf{Speech2Text}\footnote{\url{facebook/s2t-large-librispeech-asr}}~\cite{Wang2020}, \textbf{WavLM}\footnote{\url{microsoft/wavlm-large}}~\cite{Chen2022}, \textbf{w2v-BERT 2.0}\footnote{\url{facebook/w2v-bert-2.0}}~\cite{Seamless2023}, and \textbf{Wav2Vec2-Conformer}\footnote{\url{facebook/wav2vec2-conformer-rope-large}}~\cite{Wang2020}. Due to a lack of same-architecture, different-representation counterparts, commonly used models such as ECAPA-TDNN~\cite{Desplanques2020}, ReDimNet~\cite{Yakovlev2024}, and x-vectors~\cite{Snyder2018} were excluded.


All models
produced 256-dimensional speaker embeddings using temporal statistics pooling (TSTP)~\cite{Wang2023}. The ResNet34 and ResNet1D~\cite{He2016} models were trained from scratch for 150 epochs following the WeSpeaker ResNet training recipe~\cite{Wang2023}; the four pretrained models were adapted using Low-Rank Adaptation (LoRA)~\cite{Hu2022} with rank \num{16} and alpha \num{32}. All six models used the same Additive Angular Margin (AAM) Loss~\cite{Deng2019} with scale \num{32.0} and margin \num{0.2}, in which the margin is applied only once training accuracy exceeds a fixed threshold of \qty{80}{\percent}. The two convolutional models and w2v-BERT 2.0 crossed this threshold; the remaining three did not.

\subsection{Model Evaluation}
\label{subsec:model-eval}

For ASV model evaluation, we report equal error rate (EER)~\cite{Kinnunen2010} and minimum detection cost function (minDCF). Trials were scored by cosine similarity between $L^2$-normalized embeddings, without score normalization. 
We compute minDCF following~\cite{Huh2023} at
\begin{equation}
    P_{target}=0.05,\ C_{miss}=C_{fa}=1.
\end{equation}
Confidence intervals for EER values were obtained by speaker-level (cluster) paired bootstrap with \num{2000} replicates~\cite{Bolle2004}.

Baseline EER values on the unmodified signals across models differ substantially, and therefore an absolute change in EER is an imperfect metric. A system starting at a lower value has more headroom to lose, and vice versa. We therefore calculate degradation normalized by the system's headroom inspired by~\cite{Hake1998}:
\begin{equation}
\label{eq:headroom-normalized}
    D = \frac{\text{EER}_{\text{cond}} - \text{EER}_{\text{base}}}{50 - \text{EER}_{\text{base}}} \times 100,
\end{equation}
which allows us to represent the degradation as a percentage of the distance from a system's baseline to chance performance (\qty{50}{\percent}).

We report per-pair intervals and sign consistency rather than a pooled test. The nine architecture family $\times$ condition cells share models and share all trials, and therefore treating them as independent observations would be pseudoreplication~\cite{Hurlbert1984}. Significance is assessed from the bootstrap intervals rather than a parametric test. A \qty{95}{\percent} interval excluding zero is treated as evidence against a zero difference~\cite{Bengio2004}. Intervals are nominal per-contrast and are not adjusted for multiplicity.

\section{Results}
\label{sec:results}

\begin{figure}[!t]
    \centering 
    \includegraphics[width=\linewidth]{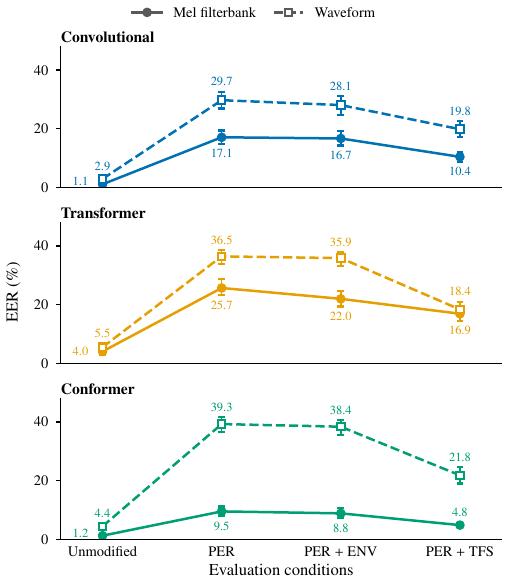}
    \caption{
    The plot visualizes EER (\%) by architecture family (Convolutional, Transformer and Conformer), evaluation condition (Unmodified, PER, PER+ENV and PER+TFS), and input representation (Mel filterbank and Waveform). Unmodified is the original VoxCeleb1-O trial audio. The conditions PER+ENV and PER+TFS are parallel restorations from PER, not successive steps. The plot shows that the performance of waveform-input models is more strongly degraded than that of filterbank-input models as temporal cues are removed. Error bars represent \qty{95}{\percent} speaker-level paired bootstrap confidence intervals over the \num{40} test speakers.
    }
    \label{fig:eer-vals}
\end{figure}

Following training, one checkpoint per model was acquired: completely trained on unmodified speech. These checkpoints were then evaluated on the degradation conditions. Results are given in \cref{tab:results} and \cref{fig:eer-vals}. 

\begin{table*}[!t]
    \centering
    
    \setlength{\tabcolsep}{4pt}
    \begin{tabular}{lccc|cc|cc|cc|cc}
    \toprule
    & & & & \multicolumn{2}{c|}{Unmodified}
    & \multicolumn{2}{c|}{PER}
    & \multicolumn{2}{c|}{PER+ENV}
    & \multicolumn{2}{c}{PER+TFS} \\
    Model & Input & Params & Pretrain
    & EER & minDCF & EER & minDCF
    & EER & minDCF & EER & minDCF \\
    \midrule
    \multicolumn{4}{@{}l}{\textit{Convolutional}} \\
    ResNet1D*   & Waveform   & 6.20M & --
    & 2.90 & 0.193 & 29.74 & \textit{0.965}
    & 28.08 & 0.930 & 19.84 & 0.802 \\
    ResNet34*   & Filterbank & 6.63M & --
    & \textbf{1.06} & \textbf{0.072} & 17.09 & 0.801
    & 16.68 & 0.785 & 10.42 & 0.579 \\
    \addlinespace
    \multicolumn{4}{@{}l}{\textit{Transformer}} \\
    WavLM       & Waveform   & 319.1M & 94k
    & 5.48 & 0.327 & 36.47 & \textit{0.971}
    & 35.86 & \textit{0.961} & 18.36 & 0.764 \\
    Speech2Text & Filterbank & 158.9M & 1k
    & 4.02 & 0.317 & 25.68 & 0.949
    & 21.98 & 0.873 & 16.89 & 0.813 \\
    \addlinespace
    \multicolumn{4}{@{}l}{\textit{Conformer}} \\
    Wav2Vec2-Conformer & Waveform & 617.6M & 960
    & 4.37 & 0.291 & 39.26 & \textit{0.993}
    & 38.36 & \textit{0.994} & 21.78 & 0.887 \\
    w2v-BERT 2.0 & Filterbank & 581.8M & 4.5M
    & 1.19 & 0.086 & \textbf{9.47} & \textbf{0.534}
    & \textbf{8.81} & \textbf{0.498} & \textbf{4.79} & \textbf{0.293} \\
    \bottomrule
    \end{tabular}
    \caption{
    Speaker verification performance on unmodified and temporally
    degraded speech, grouped by architecture family. Pretrain gives
    pretraining data in hours; the convolutional models were trained from
    scratch (*) and have none. Lower EER and minDCF indicate better
    performance; the best score in each column is bolded, and evaluations
    where minDCF exceeds \num{0.95} are italicized. Within every pair, the
    waveform-input model degrades more than its filterbank-input
    counterpart.}
    \label{tab:results}
\end{table*}

Our results are not directly comparable to~\cite{Wang2023,Wang2020,Chen2022,Seamless2023} due to differences in our training conditions.


Under decomposition most systems approach the minDCF bound of \num{1.0} as set by the minDCF operating point. Five evaluations exceed \num{0.95}, all of them obtained with waveform-input on degraded conditions. Therefore minDCF is substantially more compressed than EER in this regime and we base the input-representation contrast on EER.

\begin{figure}[!t]
    \centering
    \includegraphics[width=\linewidth]{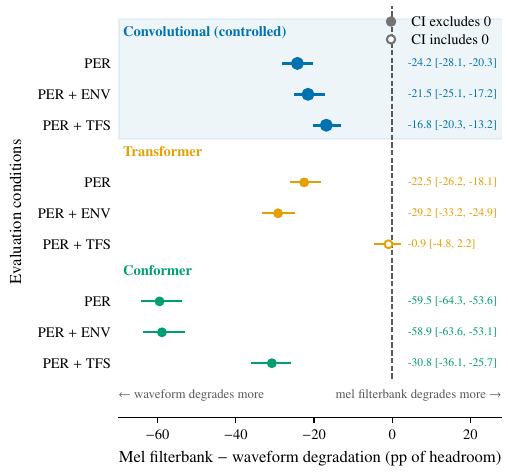}
    \caption{
    The waveform-input models degrade more in every family $\times$ condition cell. Only one of the nine contrasts has a confidence interval that includes \num{0}. Difference in headroom-normalized degradation (mel filterbank $-$ waveform) is plotted. Negative values indicate waveform models degraded more. The convolutional pair in the shaded area is the controlled comparison: identical training data and schedule, differing only in input representation. Whiskers are \qty{95}{\percent} speaker-level paired bootstrap intervals.}
    \label{fig:eer-change}
\end{figure}

The results show that all six models degrade under all three possible degradation conditions, with the PER condition causing the biggest performance drop and some models approaching \qty{40}{\percent} EER. Because EER is effectively bounded at the \qty{50}{\percent} chance level, the systems that degrade the most in the PER condition have limited room for further degradation. As such, contrasts measured at PER are conservative: ResNet1D has a larger remaining range to chance than the other two waveform-input models in this condition. Therefore the controlled comparison of the convolutional models is the least affected by the ceiling effect. PER+TFS is consistently the least damaging condition: for example w2v-BERT 2.0 reaches \qty{4.8}{\percent} EER at that condition, close to its waveform-input counterpart's unmodified performance at baseline. The ceiling compression on the PER condition does not affect this condition. The mean gain from restoring ENV, i.e. going from PER to PER+ENV, is \num{1.32} percentage points, which is dwarfed by the mean gain from restoring TFS: \num{10.94} percentage points. Performance restoration by TFS reintroduction exceeded that of ENV reintroduction in all models. The detailed degradation results can be found in \cref{tab:delta_results} and \cref{fig:eer-change}. Focusing on input representation, in all nine family $\times$ condition contrasts, the waveform-input models degrade more than the filterbank-input counterparts, though one contrast is not statistically distinguishable from zero. 

Speaker-level intervals are a median $5.97\times$ wider than trial-level intervals computed on the same scores, confirming that resampling trials individually substantially overstates precision~\cite{Bolle2004}.

\begin{table*}[!t]
    \centering
    \setlength{\tabcolsep}{3.5pt}
    \begin{tabular}{lccc|cc|cc|cc}
    \toprule
    & & & & \multicolumn{2}{c|}{PER}
    & \multicolumn{2}{c|}{PER+ENV}
    & \multicolumn{2}{c}{PER+TFS} \\
    Model & Input & Params & Pretrain
    & $\Delta$EER & $\Delta_{\mathrm{H}}$EER
    & $\Delta$EER & $\Delta_{\mathrm{H}}$EER
    & $\Delta$EER & $\Delta_{\mathrm{H}}$EER \\
    \midrule
    \multicolumn{4}{@{}l}{\textit{Convolutional}} \\
    ResNet1D*   & Waveform   & 6.20M & --
    & 26.84 & 57.0
    & 25.18 & 53.5
    & 16.94 & 36.0\\
    ResNet34*   & Filterbank & 6.63M & --
    & 16.03 & 32.8
    & 15.63 & 31.9
    & 9.36 & 19.1\\
    \addlinespace
    \multicolumn{4}{@{}l}{\textit{Transformer}} \\
    WavLM       & Waveform   & 319.1M & 94k
    & 30.99 & 69.6
    & 30.39 & 68.3
    & 12.88 & 28.9\\
    Speech2Text & Filterbank & 158.9M & 1k
    & 21.66 & 47.1
    & 17.96 & 39.1
    & 12.87 & 28.0\\
    \addlinespace
    \multicolumn{4}{@{}l}{\textit{Conformer}} \\
    Wav2Vec2-Conformer & Waveform & 617.6M & 960
    & 34.90 & 76.5
    & 34.00 & 74.5
    & 17.42 & 38.2\\
    w2v-BERT 2.0 & Filterbank & 581.8M & 4.5M
    & \textbf{8.28} & \textbf{17.0}
    & \textbf{7.62} & \textbf{15.6}
    & \textbf{3.61} & \textbf{7.4}\\
    \bottomrule
    \end{tabular}
    \caption{
    Degradation relative to unmodified speech, grouped by architecture family. $\Delta$EER is the absolute increase over each model's unmodified EER. $\Delta_{\mathrm{H}}$EER (\%) reports headroom normalized scores defined in \cref{eq:headroom-normalized}. Pretrain gives pretraining data in hours; the convolutional models were trained from scratch (*) and have none. Lower is better; the smallest degradation in each column is bolded. Within every pair, the waveform-input model degrades more than its filterbank-input counterpart.}
    \label{tab:delta_results}
\end{table*}

\section{Discussion}
\label{sec:discuss}

In every pair compared, the filterbank-input model proved to be more robust than its waveform-input counterpart for every temporal degradation. The most direct explanation of this finding is that mel filterbanks are more robust to temporal degradations than raw waveforms are. Transforms used to calculate mel filterbanks, such as magnitude-spectrum extraction, triangular frequency pooling~\cite{Davis1980}, and log compression, discard sample-level details that temporal decomposition perturbs. Mel filterbanks aim to capture higher-level characteristics of the signal and in this process of discarding sample-level information discard the perturbations introduced by the degradation. Therefore, the speaker-identifying characteristics persist. Mel filterbanks are designed to emphasize the qualities in audio that humans listen to~\cite{Stevens1937}, and therefore their use being more compatible with temporal degradation mirrors the ability of human participants to recognize speech under similar conditions~\cite{Licklider1948,Smith2002,Shannon1995}.

To strengthen this claim, we need to address the following confounds, and why they are untenable.

Pretraining scale and similarity of pretraining objective to model objective are known to improve model robustness~\cite{Hendrycks2019Pretrain}. In our experiment, WavLM received more pretraining than Speech2Text, yet the latter was more robust. A larger parameter count is also considered a contributor to robustness~\cite{Hendrycks2019Robust}, and yet in the pretrained cases the filterbank-input models were smaller and still proved to be more robust. The two convolutional models differ in the frame advance of their front ends. The mel transform advances every \qty{10}{\milli\second}, and the learned front end of the waveform-model advances at \qty{20}{\milli\second} intervals. These give effective hops of \qty{80}{\milli\second} and \qty{160}{\milli\second}, respectively, after trunk-down sampling. However, both models share an analysis window of \qty{25}{\milli\second}, meaning that they differ in sampling density. Since the other two pairs show the same pattern, frame rate alone cannot account for the effect.

To address the across-condition pattern, we need to consider that IPC lifts both noise and signal to full scale. Log compression with frequency pooling may absorb these artifacts more readily than raw waveforms might. These artifacts appear in PER and PER+TFS, but not in PER+ENV, where the envelope restores speech-silence structure. If this noise lift were the culprit, PER+ENV would have the smallest contrast at \num{-21.53}, but instead PER+TFS does at \num{-16.85} in the controlled pair, and likewise in every family.

\section{Conclusion}

We evaluated six automatic speaker verification models, across three architecture families and two input representations, on temporally degraded speech. Every system degraded under all three conditions. In each pair, the waveform-input model's performance degraded more than that of the filterbank-input counterpart. This effect is supported by the controlled convolutional pair, where all intervals exclude zero, and replicated in the two pairs of pretrained models, despite differences in their pretraining data and objective. We also show that TFS and ENV carry substantial information in speaker recognition. Under these conditions, restoring temporal fine structure recovered substantially more speaker information than restoring the envelope.


\bibliographystyle{IEEEbib}
\bibliography{refs}
\label{sec:refs}

\section{Funding Acknowledgments}
\label{sec:funding}
This project was funded by the State Secretariat for Education, Research and Innovation of the federal government of Switzerland. This work was conducted as part of the VoCS project, which has received funding from the European Union's Horizon Europe Framework program under the Marie Skłodowska-Curie Grant Agreement No 101168998.

\section{Compliance with Ethical Standards}
This study was conducted using human subject data made available in open access by VoxCeleb~\cite{Nagrani2017,Chung2018}. Ethical approval was not required as confirmed by the licensing of the data.

\end{document}